\documentclass[submission,copyright,creativecommons]{eptcs}
\providecommand{\event}{TARK 2017} % Name of the event you are submitting to
\usepackage{underscore}           % Only needed if you use pdflatex.
\usepackage{cite}

\usepackage{amsfonts}
\usepackage{amssymb}
\usepackage{amsmath}

\newtheorem{theorem}{Theorem}

\newtheorem{definition}{Definition}

\newtheorem{lemma}{Lemma}

\newtheorem{proposition}{Proposition}

\title{Rationalizability and Epistemic Priority Orderings\thanks{%
This is a short version of the paper for this volume. The full version of
the paper can be found online on the author's institutional webpage.}}
\author{Emiliano Catonini
\institute{ICEF\\
Higher School of Economics\\
Moscow, Russia}
\email{emiliano.catonini@gmail.com}
}
\date{June 2017}

\def\titlerunning{Rationalizability and Epistemic Priority Orderings}
\def\authorrunning{E. Catonini}

\begin{document}
\maketitle

\begin{abstract}
At the beginning of a dynamic game, players may have exogenous theories
about how the opponents are going to play. Suppose that these theories are
commonly known. Then, players will refine their first-order beliefs, and
challenge their own theories, through strategic reasoning. I develop and
characterize epistemically a new solution concept, Selective
Rationalizability, which accomplishes this task under the following
assumption: when the observed behavior is not compatible with the beliefs in
players' rationality and theories of all orders, players keep the orders of
belief in rationality that are per se compatible with the observed behavior,
and drop the incompatible beliefs in the theories. Thus, Selective
Rationalizability captures Common Strong Belief in Rationality (Battigalli
and Siniscalchi, 2002) and refines Extensive-Form Rationalizability (Pearce,
1984; BS, 2002), whereas Strong-$\Delta $-Rationalizability (Battigalli,
2003; Battigalli and Siniscalchi, 2003) captures the opposite epistemic
priority choice. Selective Rationalizability can be extended to encompass
richer epistemic priority orderings among different theories of opponents'
behavior. This allows to establish a surprising connection with strategic
stability (Kohlberg and Mertens, 1986).

\textbf{Keywords: }Forward induction, Strong Belief, Strong
Rationalizability, Strong-$\Delta $-Rationalizability, Strategic Stability.
\end{abstract}

\section{Introduction}

Consider the following dynamic game with perfect information.%
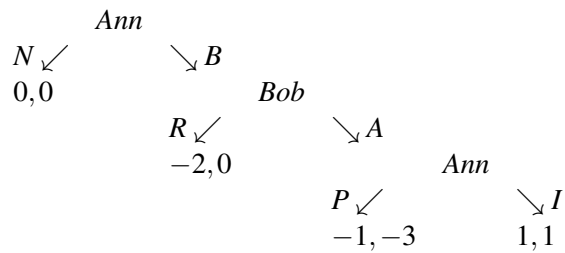
\begin{equation*}
\begin{array}{lllllll}
& Ann &  &  &  &  &  \\ 
N\swarrow &  & \searrow B &  &  &  &  \\ 
0,0 &  &  & Bob &  &  &  \\ 
&  & R\swarrow &  & \searrow A &  &  \\ 
&  & -2,0 &  &  & Ann &  \\ 
&  &  &  & P\swarrow &  & \searrow I \\ 
&  &  &  & -1,-3 &  & 1,1%
\end{array}%
\end{equation*}%
Ann can try to $B$ribe Bob, a public officer, or $N$ot. If she does, Bob can 
$A$ccept or $R$eport her, so that Ann loses two utils. If Bob accepts, Ann
can $I$mplement her plan, achieving the Pareto dominating outcome, or repent
($P$) and speak with a prosecutor, harming both Bob and herself.

\bigskip

Suppose that Ann is rational\footnote{%
i.e. subjective expected utility maximizer.} and, at the beginning of the
game, believes with probability $1$ that Bob would play $R$ after $B$. I
call this belief "(first-order belief) restriction". Then, she plays $N$.
Suppose that Bob is rational and believes that Ann is rational and that the
restriction holds. Then, he expects Ann to play $N$. So, what would Bob
believe after observing $B$? He cannot believe at the same time that Ann is
rational and that the restriction holds: the two things are at odds given $B$%
. Which of the two beliefs will Bob keep? This is the epistemic priority
issue. Suppose that he keeps the belief that the restriction holds. So, he
drops the belief that Ann is rational. Then he can also expect Ann to play $%
P $ after $(B,A)$ and so play $R$. If Ann believes that Bob reasons in this
way, she can keep her restriction and then play $N$.

\bigskip

These lines of strategic reasoning are captured by Strong-$\Delta $%
-Rationalizability (Battigalli, \cite{B2}; Battigalli and Siniscalchi, \cite%
{BS9}). In this process, the faith in the restrictions is so strong that Bob
is ready to deem Ann irrational after $B$. This could be the case if, for
instance, the belief that Bob would play $R$ is suggested by a commonly
believed social convention that always holds in context of the game (see
Battigalli and Friedenberg \cite{BF}). Suppose instead that in the context
of the game, public officers are not commonly believed to be incorruptible.
However, Bob declares that he would play $R$ in case of $B$. If Bob observes
that Ann plays $B$ anyway, he might think that Ann has not taken his words
seriously, rather than thinking that Ann is irrational. Then, Bob would
expect Ann to play $I$ after $A$, hence he would play $A$ instead of $R$. If
Ann believes that Bob is rational and keeps believing that she is rational
after $B$, she must believe that Bob will play $A$, differently than what
the restriction suggests. Hence, under this reasoning scheme, such
restriction to first-order beliefs cannot hold.

Note that opposite conclusions were reached without any uncertainty about
payoffs: the two situations do not represent two different types of Bob, but
only two different strengths of the belief that he would report Ann.

\bigskip

In Section 3, I\ construct a rationalizability procedure, Selective
Rationalizability, that captures these instances of forward induction
reasoning in dynamic games with perfect recall.\footnote{%
For notational simplicity, here the focus is kept on complete information
games. Just like Strong-$\Delta $-Rationalizability, Selective
Rationalizability can be easily extended to games with incomplete
information.} Selective Rationalizability refines a notion of Extensive-Form
Rationalizability (Pearce, \cite{P}, Battigalli and Siniscalchi, \cite{BS1}%
), which I will call "Rationalizability" for brevity. Thus, Selective\
Rationalizability represents a natural way for players to refine their
beliefs through (\emph{partial}) coordination and consequent forward
induction considerations when lone strategic reasoning about rationality
does not pin down a unique plan of actions. As above, Selective
Rationalizability delivers an empty set when the "tentative" first-order
belief restrictions of a player are at odds with strategic reasoning. In
this case, Selective Rationalizability is agnostic as to whether players
will fall back on some merely rationalizable strategy, or will still refine
their beliefs with the restrictions, up to some feasible order.

Note that strong-$\Delta $-Rationalizability, instead, does not refine
Rationalizability: in the example, $N$ is not a rationalizable outcome.%
\footnote{%
The game has no simultaneous moves and no relevant ties. Therefore, as shown
by Battigalli \cite{B97} first and Heifetz and Perea \cite{HP} later,
backward induction and Extensive-Form Rationalizability predict the same
unique outcome.} It is worth noting that Selective Rationalizability can
also be seen as an instance of Strong-$\Delta $-Rationalizability, where the
restrictions are the conjunction of the original restrictions and the
rationalizable first-order beliefs. However, keeping the two separate has
both conceptual and technical advantages. The separation allows to
investigate the epistemic priority issue between the two different sources
of belief restrictions, and to compare Strong-$\Delta $-Rationalizability
and Selective Rationalizability for the \emph{same} restrictions. In
general, one could expect Selective Rationalizability to always yield a
subset of the outcomes predicted by Strong-$\Delta $-Rationalizability. Two
counterexamples in the full version of the paper show that, (i) opposite to
the example above, Selective Rationalizability can yield non-empty
predictions when Strong-$\Delta $-Rationalizability rejects the first-order
belief restrictions; and (ii) Selective Rationalizability and Strong-$\Delta 
$-Rationalizability can even yield non-empty disjoint predictions. However,
as I show in \cite{C1}, Selective Rationalizability and Strong-$\Delta $%
-Rationalizability are outcome-equivalent when the belief restrictions
correspond to a specific path of play.

\bigskip

In Section 4, I clarify with an epistemic characterization the strategic
reasoning hypotheses that motivate Selective Rationalizability. To simplify
the epistemic analysis, the game is assumed to have a finite set of
non-terminal histories, hence finite horizon, although Selective
Rationalizability can be applied to all games with a countable set of
non-terminal histories, hence possibly infinite horizon. Selective
Rationalizability captures the behavior of rational players who restrict
their beliefs about opponents' behavior for some exogenous reason. Moreover,
at the beginning of the game, players believe that opponents are rational
and have their own restrictions; that opponents believe that everyone else
is rational and has precisely the own restrictions; and so on. These beliefs
are tentative because at some information set of a player, the observed
behavior of one opponent may be incompatible, say, with the opponent being
rational and, at the same time, having beliefs in her restricted set. In
this case, the player will drop the belief that the opponent has such
restrictions, rather than dropping the belief that the opponent is
rational.\ More generally, players always keep all orders of belief in
rationality that are per se compatible with the observed behavior, and drop
all orders of belief in the restrictions that are at odds with them. I call
this choice \emph{epistemic priority to rationality}. Strong-$\Delta $%
-Rationalizability predicts instead the behavior of players who assign
epistemic priority to the beliefs in the restrictions, and drop the
incompatible beliefs in rationality. Thus, Selective Rationalizability
captures a version of Common Strong Belief in Rationality (Battigalli and
Siniscalchi, \cite{BS1}), whereas Strong-$\Delta $-Rationalizability does
not.

\bigskip

In Section 5, I extend the analysis to finer epistemic priority orderings.
Each player can have multiple theories, say two, about opponents' behavior:
a weaker theory and a stronger theory (in the sense of more restrictive).
Players reason according to everyone's weaker theory like under Selective
Rationalizability. On top of this, as long as compatible with strategic
reasoning about the weaker theories, players reason according to the
stronger theories. So, when a player displays behavior which is not
compatible with strategic reasoning about both theories, the opponents keep
believing that the player is reasoning according to the weaker theories, and
drop the belief that the opponent is reasoning according to the stronger
ones.\footnote{%
Note that, by non-monotonicity of strong belief, strategic reasoning about
the stronger theories can potentially lead to behavior that cannot be
rationalized under the weaker theories. For this reason, the epistemic
priority issue arises.} In this short version of the paper, I consider two
theories that correspond to an equilibrium path and an equilibrium strategy
profile. This allows to establish a surprising connection with strategic
stability (Kohlberg and Mertens \cite{KM}).

\bigskip 

Since players' theories of opponents' behavior are assumed to be commonly
known,\footnote{%
In the sense that players know what theories the opponents are supposed to
have.} the most natural application of Selective Rationalizability is
explicit, pre-play coordination among players.\ Since a non-binding
agreement is purely cheap talk, if a player displays behavior which is not
compatible with rationality and belief in the agreement, the opponents are,
in my view, more likely to abandon the belief that the player believes in
the agreement, rather than the belief that the opponent is rational. As in
the example, the agreement can also be interpreted as a set of public
announcements.\footnote{%
Or, extending Selective Rationalizability to incomplete information games,
the restrictions can model public news about a state of nature. For
instance, in a financial market, players can tentatively believe that
everyone is reasoning according to the same public information about a state
of nature. Yet, if a player does not behave accordingly, the opponents may
believe that the player has different information rather than deeming the
player irrational.} Thus, Selective Rationalizability seems to be an
appropriate tool to combine strategic reasoning and equilibrium play,
especially when the motivation for equilibrium is explicit coordination. The
application of Selective Rationalizability to agreements and its
relationship with equilibrium are deeply investigated in \cite{C}. In
particular, the outcomes that Selective Rationalizability uniquely pins down
for some restrictions do not include and are not included in the set of
subgame perfect equilibrium outcomes. However, I show in \cite{C9} that
there always exists a subgame perfect equilibrium in behavioral strategies
whose possible outcomes are delivered by Selective Rationalizability for
particular restrictions. It is worth noting that the flexibility of
Selective Rationalizability, which allows to model incomplete coordination
instead of coordination on full strategy profiles, can be crucial to induce
an outcome of the game (see \cite{C} for details).

\bigskip

\section{Preliminaries}

Consider a finite dynamic game with complete information and perfect recall.
Some notation:

\begin{itemize}
\item $I$ is the finite set of \emph{players}, and for any \emph{profile} $%
(X_{i})_{i\in I}$ and any $\emptyset \not=J\subseteq I$, I write $%
X_{J}:=\times _{j\in J}X_{j}$, $X:=X_{I}$, $X_{-i}:=X_{I\backslash \left\{
i\right\} }$, $X_{-i,j}:=X_{I\backslash \left\{ i,j\right\} }$;

\item $H_{i}$ is the set of \emph{information sets} of player $i$, endowed
with the precedence relation $\prec $;

\item $Z$ is the set of\emph{\ terminal histories}$;$

\item $u_{i}:Z\rightarrow 
%TCIMACRO{\U{211d} }%
%BeginExpansion
\mathbb{R}
%EndExpansion
$ is the \emph{payoff function} of player $i$.
\end{itemize}

A\emph{\ strategy} is a function $s_{i}:h\in H_{i}\mapsto s_{i}(h)\in
A_{i}(h)$, where $A_{i}(h)$ is the set of available actions of player $i$ at
information set $h$. The set of all strategies is denoted by $S_{i}$. A
strategy profile clearly induces one and only one terminal history; let $%
\zeta :S\rightarrow Z$ denote the map which associates each strategy profile 
$s\in S$ with the induced terminal history $z\in Z$. The set of strategies
of player $i$ which allow to reach an information set $h$ (not necessarily
of player $i$!) is%
\begin{equation*}
S_{i}(h):=\left\{ s_{i}\in S_{i}:\exists s_{-i}\in S_{-i},\exists x\in
h,x\prec \zeta ((s_{i},s_{-i}))\right\} .
\end{equation*}%
For any $(\overline{S}_{j})_{j\in I}\subset S$, let $\overline{S}%
_{i}(h):=S_{i}(h)\cap \overline{S}_{i}$. If $h\in H_{i}$, $S_{-i}(h)$
represents the partial observation by player $i$ of opponents' strategies up
to $h$. For any $J\subseteq I$, $H_{i}(\overline{S}_{J}):=\left\{ h\in H_{i}:%
\overline{S}_{J}(h)\not=\emptyset \right\} $ is the set of information sets
of $i$ compatible with $\overline{S}_{J}$.

\bigskip

Players update their beliefs about opponents' strategies and beliefs as the
game unfolds. A Conditional Probability System (Renyi, \cite{R}; henceforth
CPS) assigns to each information set a belief, conditional on the observed
opponents' behavior. Here I define CPS's over the opponents' state space $%
\Omega _{-i}:=\times _{j\not=i}(S_{j}\times T_{j})$, where epistemic type
spaces $(T_{j})_{j\in I}$ will be defined in Section 4.

\begin{definition}
A Conditional Probability System on $(\Omega _{-i},(T_{-i}\times
S_{-i}(h))_{h\in H_{i}})$, with Borel sigma algebra $\boldsymbol{B}(\Omega
_{-i})$, is a mapping $\mu (\cdot |\cdot ):\boldsymbol{B}(\Omega
_{-i})\times (T_{-i}\times S_{-i}(h))_{h\in H_{i}}\rightarrow \lbrack 0,1]$
satisfying the following axioms:

\begin{enumerate}
\item[CPS-1.] for every $C\in (T_{-i}\times S_{-i}(h))_{h\in H_{i}}$, $\mu
(C|C)=1$;

\item[CPS-2.] for every $C\in (T_{-i}\times S_{-i}(h))_{h\in H_{i}}$, $\mu
(\cdot |C)$ is a probability measure on $\Omega _{-i}$;

\item[CPS-3.] for every $E\in \boldsymbol{B}(\Omega _{-i})$ and $B,C\in
(T_{-i}\times S_{-i}(h))_{h\in H_{i}}$, if $E\subseteq B\subseteq C$ then $%
\mu (E|B)\mu (B|C)=\mu (E|C)$.
\end{enumerate}
\end{definition}

The set of all CPS's of player $i$ is denoted by $\Delta ^{H_{i}}(\Omega
_{-i})$.\footnote{%
If each $\Omega _{i}$ is compact metrizable, endowing the set $\Delta
(\Omega _{-i})$ of Borel probability measures on $\Omega _{-i}$ with the
topology of weak convergence and $(\Delta (\Omega _{-i}))^{H_{i}}$ with the
product topology, Battigalli and Siniscalchi \cite{BS0} prove that $\Delta
^{H_{i}}(\Omega _{-i})$ is a compact metrizable subset of $(\Delta (\Omega
_{-i}))^{H_{i}}$.} For brevity, conditioning events will be indicated with
just the information set.

CPS's on strategies are defined by replacing $\Omega _{-i}$ with $S_{-i}$
and $(T_{-i}\times S_{-i}(h))_{h\in H_{i}}$ with $(S_{-i}(h))_{h\in H_{i}}$.
For any $J\subseteq I\backslash \left\{ i\right\} $ and $\overline{S}%
_{J}\subseteq S_{J}$, I say that $\mu _{i}\in \Delta ^{H_{i}}(S_{-i})$ \emph{%
strongly believes} (Battigalli and Siniscalchi, \cite{BS1})\footnote{%
Battigalli and Siniscalchi make a stricter use of the term strong belief, by
referring only to Borel subsets of $\Omega _{-i}$ or $S_{-i}$.} $\overline{S}%
_{J}$ if $\mu _{i}(\overline{S}_{J}\times S_{I\backslash (J\cup \left\{
i\right\} )}|h)=1$ for all $h\in H_{i}(\overline{S}_{J})$.

\bigskip

I\ consider players who reply rationally to their conjectures. By
rationality I\ mean that players, at every information set, choose an action
that maximizes expected utility given their belief about how the opponents
will play and the expectation to choose rationally again in the continuation
of the game. This is equivalent (see Battigalli \cite{B3}) to playing a 
\emph{sequential best reply} to the CPS.

\begin{definition}
Fix $\mu _{i}\in \Delta ^{H_{i}}(S_{-i})$. A strategy $s_{i}\in S_{i}$ is a 
\textsl{sequential best reply} to $\mu _{i}$ if for each $h\in H_{i}(s_{i})$%
, $s_{i}$ is a \emph{continuation best reply} to $\mu _{i}(\cdot |h)$, i.e.
for all $\widetilde{s}_{i}\in S_{i}(h)$,%
\begin{equation*}
\sum \limits_{s_{-i}\in S_{-i}(h)}u_{i}(\zeta (s_{i},s_{-i}))\mu
_{i}(s_{-i}|h)\geq \sum \limits_{s_{-i}\in S_{-i}(h)}u_{i}(\zeta (\widetilde{%
s}_{i},s_{-i}))\mu _{i}(s_{-i}|h)\text{.}
\end{equation*}
\end{definition}

The set of sequential best replies to a CPS $\mu _{i}\in \Delta
^{H_{i}}(S_{-i})$ is denoted by $\rho (\mu _{i})$.

\bigskip

\section{Selective Rationalizability}

Before defining Selective Rationalizability, I\ have to pin down the
behavior of players when they only reason about rationality. This task has
already been accomplished in the literature under different assumptions.
Pearce \cite{P} defines Extensive-Form Rationalizability under \emph{%
structural consistency} (an underlying feature also of sequential
equilibrium). Battigalli \cite{B1} assumes \emph{strategic independence},
which requires players to maintain the first-order belief about each
opponent whenever her individual behavior does not contradict them.
Battigalli and Siniscalchi \cite{BS1} remove any assumption of independence
and require players to maintain each order of belief in rationality only
until \emph{none} of the opponents contradict it. Then, they give to the
resulting elimination procedure, \emph{Strong Rationalizability}, an
epistemic characterization based on the notion of \emph{strong belief}. For
this reason, I adopt Strong Rationalizability as a starting point, but I
amend it by introducing \emph{independent rationalization}: players maintain
an order of belief in rationality of an opponent as long as her individual
behavior does not contradict it. The motivation for this choice is two-fold.
First, it is coherent with the emphasis on the persistence of beliefs in
rationality. Second, there is an important motivation for the adoption of
independent rationalization in Selective Rationalizability, which will be
explained later. As far as Strong Rationalizability is concerned, it is easy
to observe that independent rationalization is immaterial for the predicted
outcomes, since it kicks in at an information set only when it is not
reached anymore by some opponent. Instead, I do not adopt strategic
independence. This is not in contradiction with independent rationalization:
there can be correlations\footnote{%
For instance, a player can believe that a sunny day will induce more
optimistic beliefs in two opponents (regardless of their strategic
sophistication); see, for instance, Aumann \cite{A} and Brandenburger and
Friedenberg \cite{BRF}.} also among the choices of players with different
orders of belief in rationality (and actually, players do commonly believe
in rationality along the rationalizable paths). However, assuming strategic
independence would complicate the notation but not alter the results. For
brevity and to distinguish it from the original notion of Strong
Rationalizability, I will call this version simply "Rationalizability".

\begin{definition}[Rationalizability]
Consider the following procedure.\medskip

\emph{(Step 0)} For each $i\in I$, let $S_{i}^{0}=S_{i}$.\medskip

\emph{(Step n$>$0)} For each $i\in I$ and $s_{i}\in S_{i}$, let $%
s_{i}\in S_{i}^{n}$ if and only if there is $\mu _{i}\in \Delta
^{H_{i}}(S_{-i})$ such that:

\begin{enumerate}
\item[R1] $s_{i}\in \rho (\mu _{i})$;

\item[R2] $\mu _{i}$ strongly believes $S_{j}^{q}$ for all $j\not=i$ and $%
q<n $.\medskip
\end{enumerate}

Finally let $S_{i}^{\infty }=\cap _{n\geq 0}S_{i}^{n}$. The profiles in $%
S^{\infty }$ are called rationalizable.
\end{definition}

Strong-$\Delta $-Rationalizability is defined exactly like Strong
Rationalizability, except that at each step only beliefs $\mu _{i}$ in a
restricted set of CPS's $\Delta _{i}\subset \Delta ^{H_{i}}(S_{-i})$ are
allowed.

\bigskip

Selective Rationalizability refines Rationalizability in the following way.
Each player has an exogenous theory of opponents' behavior and refines the
rationalizable first-order beliefs according to this theory. The theory of
player $i$ is represented by a set of CPS's $\Delta _{i}\subseteq \Delta
^{H_{i}}(S_{-i})$ over opponents' strategies. Players are aware of the
theories of everyone else. Therefore, players can also expect \emph{each}
opponent to refine her first-order beliefs according to the own theory. This
expectation towards an opponent is maintained as long as the opponent
herself is not observed making a move that contradicts it. Moreover, players
expect \emph{each} opponent to reason about everyone else in the same way.
Also this expectation is maintained as long as the opponent herself does not
make a move that contradicts it. And so on. Thus, Selective
Rationalizability is defined under independent rationalization. This allows
better comparability with the equilibrium literature. Without independent
rationalization, if a player deviates from the agreed-upon equilibrium path,
each opponent is free to believe that any other opponent is not going to
implement her threat. In this way, no coordination of threats would be
required. These issues are widely discussed in \cite{C}. Note however that
independent rationalization is immaterial for the message of this paper and
for the analysis of all the examples: players are only two in all games
except for the game of Section 5, where independent rationalization plays no
role anyway.

\begin{definition}[Selective Rationalizability]
Fix a profile $(\Delta _{i})_{i\in I}$ of compact subsets of CPS's.\\ Let $%
((S_{i}^{m})_{i\in I})_{m=0}^{\infty }$ denote the Rationalizability
procedure. Consider the following procedure.\medskip

\emph{(Step 0)} For each $i\in I$, let $S_{i,R\Delta }^{0}=S_{i}^{\infty }$%
.\medskip

\emph{(Step n$>$0)} For each $i\in I$ and $s_{i}\in S_{i}$, let $%
s_{i}\in S_{i,R\Delta }^{n}$ if and only if there is $\mu _{i}\in \Delta
_{i} $ such that:

\begin{enumerate}
\item[S1] $s_{i}\in \rho (\mu _{i})$;

\item[S2] $\mu _{i}$ strongly believes $S_{j,R\Delta }^{q}$ for all $j\not=i$
and $q<n$;

\item[S3] $\mu _{i}$ strongly believes $S_{j}^{q}$ for all $j\not=i$ and $%
q\in 
%TCIMACRO{\U{2115} }%
%BeginExpansion
\mathbb{N}
%EndExpansion
$.\medskip
\end{enumerate}

Finally, let $S_{i,R\Delta }^{\infty }=\cap _{n\geq 0}S_{i,R\Delta }^{n}$.
The profiles in $S_{R\Delta }^{\infty }$ are called
selectively-rationalizable.
\end{definition}

Step 0 initializes Selective\ Rationalizability with the rationalizable
strategy profiles. This is only to stress that Selective\ Rationalizability
refines Rationalizability: S3 already implies that players strongly believe
in the rationalizable strategies of each opponent, and that the strategies
surviving Step 1 are rationalizable. Indeed, Selective Rationalizability can
also be seen as an extension of Rationalizability, in a unique elimination
procedure where the first-order belief restrictions kick in once no more
strategies can be eliminated otherwise.

\bigskip

Selective Rationalizability can be simplified in different ways according to
the structure of the restrictions. S3 can be substituted by the requirement
that strategies be rationalizable when first-order beliefs are not
restricted at the non-rationalizable information sets.

\begin{definition}
\label{RAT RESTR}I$\ $say that $\Delta _{i}\subseteq \Delta ^{H_{i}}(S_{-i})$
is rationalizable if $\mu _{i}^{\ast }\in \Delta _{i}$ whenever there exists 
$\mu _{i}\in \Delta _{i}$ such that $\mu _{i}^{\ast }(\cdot |h)=\mu
_{i}(\cdot |h)$ for all $h\in H_{i}(S^{\infty })$.
\end{definition}

\begin{proposition}
\label{NO S3}Suppose that for every $i\in I$, $\Delta _{i}$ is
rationalizable. Then, S3 can be substituted by $s_{i}\in
S_{i}^{0}=S_{i}^{\infty }$ in the definition of Selective\ Rationalizability.
\end{proposition}

\begin{proposition}
\label{NO NEED RESTRICT}Fix $(\Delta _{i})_{i\in I}\subseteq \times _{i\in
I}\Delta ^{H_{i}}(S_{-i})$ with $S_{R\Delta }^{\infty }\not=\emptyset $.
There exists a profile $(\Delta _{i}^{\ast })_{i\in I}$ of rationalizable%
\footnote{%
Although it is not formally shown, it is straightforward to observe that if
each $\Delta _{i}$ is compact, each $\Delta _{i}^{\ast }$ constructed in the
proof is compact too. This shows that the epistemic characterization of
Selective Rationalizability holds under $(\Delta _{i}^{\ast })_{i\in I}$; it
is however immaterial for the results of this section.} subsets of CPS's
such that $\zeta (S_{R\Delta ^{\ast }}^{\infty })=\zeta (S_{R\Delta
}^{\infty })$.
\end{proposition}

Thus, the class of rationalizable restrictions suffices to yield all the
possible behavioral implications of Selective Rationalizability.

\bigskip

Selective Rationalizability and Strong-$\Delta $-Rationalizability can yield
the empty set. This happens when at some step there is no $\mu _{i}\in
\Delta _{i}$ that satisfies S2 and S3, or the equivalent of S2 for Strong-$%
\Delta $-Rationalizability. This means that the restrictions are not
compatible with strategic reasoning about rationality and the restrictions
themselves.

\bigskip

\section{Epistemic framework and characterization theorem}

I\ adopt the epistemic framework of Battigalli and Prestipino \cite{BP},
dropping the incompleteness of information dimension. Players' beliefs over
strategies of all orders are given an implicit representation through a
compact, complete and continuous \emph{type structure} $(\Omega
_{i},T_{i},g_{i})_{i\in I}$,\footnote{%
Friedenberg \cite{F0} proves that in static games, such a type structure
contains all hierarchies of beliefs about strategies. Although this result
has not been formally extended to dynamic games, to the best of my
knowledge, no counterexample to this extension has ever been found. However,
the canonical type structure for CPS of Battigalli and Siniscalchi \cite{BS0}
is compact, complete, and continuous (and it contains all collectively
coherent hierarchies of beliefs by construction).} where for every $i\in I$, 
$\Omega _{i}=S_{i}\times T_{i}$, $T_{i}$ is a compact metrizable space of 
\emph{epistemic types}, and $g_{i}=(g_{i,h})_{h\in H_{i}}:T_{i}\rightarrow
\Delta ^{H_{i}}(\Omega _{-i})$ is a continuous and onto\footnote{%
This imposes to choose type spaces with the cardinality of the continuum.} 
\emph{belief map}. I will call "events" the elements of the Borel
sigma-algebras on each $\Omega _{i}$, and of the product sigma algebras on
the Cartesian spaces $\Omega _{J}:=\times _{i\in J\subseteq I}\Omega _{i}$.

\hspace{-2mm}%
The first-order belief map of player $i$, $f_{i}=(f_{i,h})_{h\in
H_{i}}\!:T_{i}\rightarrow \Delta ^{H_{i}}(S_{-i})$, is defined as $%
f_{i,h}(t_{i})=$\textit{$\mathrm{Marg}$}$_{S_{-i}}g_{i,h}(t_{i})$ for all $%
i\in I$ and $h\in H_{i}$, so it inherits continuity from $g_{i}$. The event
in $\Omega _{i}$ where the restrictions of player $i$ hold is%
\begin{equation*}
\lbrack \Delta _{i}]:=\left \{ (s_{i},t_{i})\in \Omega _{i}:f_{i}(t_{i})\in
\Delta _{i}\right \} ;
\end{equation*}%
$[\Delta _{i}]$ is compact because $\Delta _{i}$ is compact and $f_{i}$ is
continuous. The cartesian set where the restrictions of all players hold is $%
[\Delta ]:=\times _{i\in I}[\Delta _{i}]$.

From now on, fix a Cartesian (across players) event $E=\times _{i\in
I}E_{i}\subseteq \Omega $. The closed\footnote{%
See Battigalli and Prestipino \cite{BP}.} event where player $i$ believes in 
$E_{-i}$ at an information set $h\in H_{i}$ is defined as%
\begin{equation*}
B_{i,h}(E_{-i}):=\left\{ (s_{i},t_{i})\in \Omega
_{i}:g_{i,h}(t_{i})(E_{-i})=1\right\} .
\end{equation*}%
The closedness of $B_{i,h}(E_{-i})$ implies the closedness of all the
following belief events. The event where $i$ believes in $E_{-i}$ at every
information set is $B_{i}(E_{-i}):=\cap _{h\in H_{i}}B_{i,h}(E_{-i})$.

If \emph{$\mathrm{Proj}$}$_{S}E=S$, $E$ is an \emph{epistemic event}. Else,
it could be impossible for player $i$ to believe in $E_{-i}$ at some
information set $h\in H_{i}$, because \emph{$\mathrm{Proj}$}$%
_{S_{-i}}E_{-i}\cap S_{-i}(h)=\emptyset $. However, player $i$ may want to
believe in $E_{-i}$ as long as not contradicted by observation. The event
where this persistency of the belief holds is:%
\begin{equation*}
\overline{SB}_{i}(E_{-i}):=\bigcap\limits_{h\in H_{i}:\mathrm{Proj}%
_{S_{-i}}E_{-i}\cap S_{-i}(h)\not=\emptyset }B_{i,h}(E_{-i}).
\end{equation*}%
The \emph{strong belief} operator $\overline{SB}_{i}$ is non-monotonic: if $%
E_{-i}\subset F_{-i}$, it needs not be the case that $\overline{SB}%
_{i}(E_{-i})\subset \overline{SB}_{i}(F_{-i})$. This will explain why Strong-%
$\Delta $-Rationalizability is not a refinement of Strong Rationalizability,
and Selective Rationalizability, for the same restrictions, is not a
refinement of Strong-$\Delta $-Rationalizability.

Suppose now that, for each opponent $j$, player $i$ believes that the true
pair $(s_{j},t_{j})$ is in $E_{j}$, as long as this is not contradicted by
observation. Then I say that $i$ strongly believes in $E_{j}$ for all $%
j\not=i$. Formally, I define the \emph{independent strong belief} operator as%
\begin{equation*}
SB_{i}(E_{-i}):=\cap _{j\not=i}\overline{SB}_{i}(E_{j}\times \Omega _{-j,i}).
\end{equation*}%
Note that $(s_{i},t_{i})\in SB_{i}(E_{-i})$ if and only if, for each $%
j\not=i $, $g_{i}(t_{i})$ strongly believes in $E_{j}$, i.e. $%
g_{i,h}(t_{i})(E_{j}\times \Omega _{-i,j})=1$ for all $h\in H_{i}$ with $%
\mathrm{Proj}_{S_{j}}E_{j}\cap S_{j}(h)\not=\emptyset $.

Let $B(E):=\times _{i\in I}B_{i}(E_{-i})$, $SB(E):=\times _{i\in
I}SB_{i}(E_{-i})$, and $CSB_{i}(E):=E_{i}\cap SB_{i}(E_{-i})$. The correct
and mutual strong belief in the event $E$ is denoted by:%
\begin{equation*}
CSB(E):=\times _{i\in I}CSB_{i}(E)=E\cap SB(E).
\end{equation*}%
Let $B^{0}(E):=E=CSB^{0}(E)$. For all $n\in 
%TCIMACRO{\U{2115} }%
%BeginExpansion
\mathbb{N}
%EndExpansion
$, define the following $n$-th order belief operators: $%
B^{n+1}(E):=B(B^{n}(E))$ and $CSB^{n+1}(E):=CSB(CSB^{n}(E))$. An epistemic
event $E$ is \emph{transparent} when it holds and is believed by all players
at every information set and at every order. The corresponding event is $%
B^{\ast }(E):=\cap _{n\geq 0}B^{n}(E)$. If $E$ is not an epistemic event, I
will be interested in the event $CSB^{\infty }(E):=\cap _{n\geq 0}CSB^{n}(E)$%
.

First-order beliefs and higher-order beliefs have no bite in terms of
behavior and predictions over opponents' behavior without rationality and
beliefs in rationality. The "rationality of player $i$" event is denoted by%
\begin{equation*}
R_{i}:=\left\{ (s_{i},t)\in \Omega _{i}:s_{i}\in \rho (f_{i}(t_{i}))\right\}
,
\end{equation*}%
and it is closed whenever $\rho \circ f_{i}$, as assumed here, is
upper-hemicontinuous.\footnote{%
Finiteness suffices for upper-hemicontinuity.} The \emph{rationality} event
is $R:=\times _{i\in I}R_{i}$.

\bigskip

Here I consider rational players who keep, as the game unfolds, the highest
order of belief in rationality of each opponent that is consistent with her
observed behavior. Players further refine their first-order beliefs through
the own restrictions. All this is captured by the event $\left[ \Delta %
\right] \cap CSB^{\infty }(R)$. The event "rationality and common
independent strong belief in rationality" $CSB^{\infty }(R)$ characterizes
Rationalizability.\footnote{%
Analogously, "rationality and common strong belief in rationality"
characterizes Strong Rationalizability (Battigalli and Siniscalchi \cite{BS1}%
).} Furthermore, players believe, as long as not contradicted by
observation, that each opponent: (1) reasons in the same way; (2) believes,
as long as not contradicted by observation, that everyone else reasons in
the same way; and so on. The $n$-th order of this belief is captured by the
event $CSB^{n}(\left[ \Delta \right] \cap CSB^{\infty }(R))$, and it
characterizes the $n+1$-th step of Selective\ Rationalizability. The event $%
CSB^{\infty }(\left[ \Delta \right] \cap CSB^{\infty }(R))$ captures all the
steps of reasoning at once.

\begin{theorem}
\label{T4}Fix a profile $\Delta =(\Delta _{i})_{i\in I}$ of compact subsets
of CPS's. Then, for every $n\geq 0$,%
\begin{equation*}
S_{R\Delta }^{n+1}=\mathrm{Proj}_{S}CSB^{n}(\left[ \Delta \right] \cap
CSB^{\infty }(R)),
\end{equation*}%
and 
\begin{equation*}
S_{R\Delta }^{\infty }=\mathrm{Proj}_{S}CSB^{\infty }(\left[ \Delta \right]
\cap CSB^{\infty }(R)).
\end{equation*}
\end{theorem}

\bigskip

The comparison between the characterization of Selective Rationalizability
and the characterization of Strong-$\Delta $-Rationalizability proposed by
Battigalli and Prestipino \cite{BP} clarifies the epistemic priority
difference behind the two solution concepts. In the event $\overline{CSB}%
^{\infty }(R\cap B^{\ast }([\Delta ]))\subset B^{\ast }([\Delta ])$ that
characterizes Strong-$\Delta $-Rationalizability,\footnote{$\overline{CSB}%
^{\infty }$ is defined like $CSB^{\infty }$ starting from $\overline{SB}_{i}$
instead of $SB_{i}$.} players keep at every information set every order of
belief in the restrictions. In the event $CSB^{\infty }(\left[ \Delta \right]
\cap CSB^{\infty }(R))\subset CSB^{\infty }(R)$, players keep at every
information set the highest order of belief in each opponent's rationality
which is per se compatible with her observed behavior.

\bigskip

\section{Finer epistemic priority orderings}

Consider the following game, where after $I$ Cleo chooses the matrix.%
\begin{equation*}
\begin{tabular}{cccc|c|c|c|}
\cline{5-7}
&  &  &  & $M1$ & $L$ & $R$ \\ \cline{5-7}
& $Cleo$ & --- $\ I\longrightarrow $ &  & $U$ & $1,1,3.3$ & $0,0,3.3$ \\ 
\cline{5-7}
& $\downarrow O$ &  &  & $D$ & $0,0,3.3$ & $1,1,3.9$ \\ 
\cline{1-3}\cline{5-7}\cline{5-7}
\multicolumn{1}{|c}{$A\backslash B$} & \multicolumn{1}{|c}{$W$} & 
\multicolumn{1}{|c}{$E$} & \multicolumn{1}{|c|}{} & $M2$ & $L$ & $R$ \\ 
\cline{1-3}\cline{5-7}
\multicolumn{1}{|c}{$N$} & \multicolumn{1}{|c}{$2,2,3.6$} & 
\multicolumn{1}{|c}{$0,0,0$} & \multicolumn{1}{|c|}{} & $U$ & $0,0,0$ & $%
1,1,8.1$ \\ \cline{1-3}\cline{5-7}
\multicolumn{1}{|c}{$S$} & \multicolumn{1}{|c}{$0,0,0$} & 
\multicolumn{1}{|c}{$2,2,4$} & \multicolumn{1}{|c|}{} & $D$ & $1,1,8.1$ & $%
0,0,0$ \\ \cline{1-3}\cline{5-7}
\end{tabular}%
\end{equation*}%
All strategies are rationalizable. Suppose that players have theories of
opponents' behavior that come from an equilibrium or an incomplete agreement
among all players. An incomplete agreement or an equilibrium\footnote{%
With the notable exception of self-confirming equilibrium (Fudenberg and
Levine \cite{FL}).} align any two players' beliefs about a third player's
moves. Are there restrictions of this kind under which Selective
Rationalizability yields outcome $(O,(S,E))$? Yes. It is sufficient that
Cleo expects Ann and Bob to play $(S,E)$ after $O$ and, for instance, $(U,L)$
after $I$. Then, upon observing $I$, Ann and Bob drop the belief that Cleo
has the aforementioned first-order belief restrictions. Thus, they can
expect Cleo to pick any of the two matrices. If they believe that Cleo picks
matrix $M1$, Ann may play $U$ when she believes that Bob will play $L$, and
vice versa.

Suppose now instead that Ann and Bob have an alternative theory to
rationalize Cleo's move. They believe that Cleo believed that they would
have coordinated on $(S,E)$ after $O$, but does not believe that they will
play $(U,L)$ after $I$. If Ann and Bob rationalize the move of Cleo under
this light, they expect Cleo to pick $M2$, because $(I,M1)$ is not rational
given the belief in $(S,E)$. Under $M2$, Ann and Bob cannot coordinate on $%
(U,L)$.

Suppose now that Cleo expects Ann and Bob to play $(N,W)$ after $O$ and $%
(U,L)$ after $I$. Upon observing $I$, as above, Ann and Bob believe the Cleo
believed in $(N,W)$ after $O$, but does not believe in $(U,L)$ after $I$.
But this does not exclude that Cleo would play $M1$, hoping in $(D,R)$.
Thus, Ann may play $U$ when she believes that Bob will play $L$, and vice
versa. So, Cleo's initial restrictions are compatible with the belief that
Ann and Bob have the same restrictions about each other's moves, and will
rationalize $I$ under her belief in $(N,W)$ only. The restrictions yield
outcome $(O,(N,W))$ as unique prediction not just under Selective
Rationalizability, but also under the additional strategic reasoning
hypotheses.

Note a paradoxical fact: to convince Cleo to play $O$, Ann and Bob must
promise to play $(N,W)$, which yields Cleo a payoff of $3.6$, instead of $%
(S,E)$, which yields Cleo a payoff of $4$. The intuitive explanation is that
a higher expectation of Cleo after $O$ allows her to take a convincing
position of power after $I$.

\bigskip

Two important questions arise now. First: does the exclusion of $(O,(S,E))$
and not of $(O,(N,W))$ correspond to some existing equilibrium refinement?
Note that both outcomes are induced by a subgame perfect equilibrium in
(extensive-form/strongly) rationalizable strategies. Second, and most
importantly: can the strategic reasoning above be modeled as an epistemic
priority order between different theories of opponents behavior and be
captured by a solution concept analogous to Selective Rationalizability?

The answer to the first question is yes: strategic stability a la Kohlberg
and Mertens \cite{KM}.\footnote{%
Strategic stability has been chosen over Forward Induction equilibria of
Govindan and Wilson \cite{G} or Man \cite{M} because the latter do not
refine extensive-form rationalizability, hence do not capture all orders of
strong belief in rationality. Strategic stability, instead, refines iterated
admissibility, which in generic games corresponds to extensive-form
rationalizability (Shimoji, \cite{SH}).}

\begin{definition}[Kohlberg and Mertens \protect\cite{KM}]
\label{KM}For each $i\in I$, let $\Sigma _{i}$ be the set of mixed
strategies of $i$, i.e. the set of probability distributions over $S_{i}$. A
closed set of mixed equilibria $\widehat{\Sigma }\subseteq \Sigma $ is
stable if it is minimal with respect to the following property: for any $%
\varepsilon >0$, there exists $\delta _{0}>0$ such that for any completely
mixed $(\sigma _{i})_{i\in I}\in \Sigma $ and $(\delta _{i})_{i\in I}$ with $%
\delta _{i}<\delta _{0}$ for all $i\in I$, the perturbed game where for
every $i\in I$, every $s_{i}\in S_{i}$ is substituted by $(1-\delta
_{i})s_{i}+\delta _{i}\sigma _{i}$ has a mixed equilibrium $\varepsilon $%
-close to $\widehat{\Sigma }$.
\end{definition}

Consider first a set of two mixed equilibria $\widehat{\Sigma }=\left\{
(\sigma _{i})_{i\in I},(\sigma _{i}^{\prime })_{i\in I}\right\} $ inducing
outcome $(O,(N,W))$, where $\sigma _{C}(O)=\sigma _{C}^{\prime }(O)=1$, $%
\sigma _{A}(N.D)=\sigma _{B}(W.R)=1/\sqrt{2}$, and $\sigma _{A}^{\prime
}(N.D)=\sigma _{B}^{\prime }(W.R)=2/3$. Under $\sigma $, Cleo is actually
indifferent between $O$ and $I.M1$, while under $\sigma ^{\prime }$, she is
indifferent between $O$ and $I.M2$. I show that $\widehat{\Sigma }$ is
stable. Fix any completely mixed $(\widetilde{\sigma }_{i})_{i\in I}\in
\Sigma $, an arbitrarily small $\delta _{0}$, and $(\delta _{i})_{i\in I}$
with $\delta _{i}<\delta _{0}$ for all $i\in I$. Consider the game perturbed
as in Definition \ref{KM} and indicate with tilde the perturbed strategies.
If $\widetilde{\sigma }_{A}(I.M1)>\widetilde{\sigma }_{A}(I.M2)$ (resp., $%
\widetilde{\sigma }_{A}(I.M1)<\widetilde{\sigma }_{A}(I.M2)$), assign small
probability to $\widetilde{I.M1}$ (resp., $\widetilde{I.M2}$) and the
complementary probability to $\widetilde{O}$ in such a way that $I.M1$ and $%
I.M2$ are played with probability $1/2$. Then, after $I$, Ann and Bob are
indifferent between their actions regardless of the belief about the action
of the other. Thus, since all strategies are perturbed in the same way, Ann
and Bob are indifferent between $\widetilde{N.U}$ and $\widetilde{N.D}$, and
between $\widetilde{W.L}$ and $\widetilde{W.R}$. Assign probability to these
strategies in such a way that Cleo is indifferent between $\widetilde{O}$
and $\widetilde{I.M1}$ (resp., $\widetilde{I.M2}$).\footnote{%
Since the perturbed strategies assign positive probability to $S$ and $E$,
the expected payoff of Cleo after $O$ is lower than $3.6$. Thus, $N.U$ and $%
N.L$ (resp., $N.D$ and $N.R$) must be assigned probability\ higher than $1/%
\sqrt{2}$ (resp., than $2/3$).} For any $\varepsilon >0$, by picking a small
enough $\delta _{0}$, we have an equilibrium in the perturbed strategies
where the induced probabilities over the original strategies are $%
\varepsilon $-close to those assigned by $\sigma $ (resp., $\sigma ^{\prime
} $).

Instead, there is no stable set of equilibria inducing $(O,(S,E))$: any
perturbation of $O$ that gives negligible probability to $I.M1$ with respect
to $I.M2$ cannot be compensated by giving positive probability to $%
\widetilde{I.M1}$, because $\widetilde{I.M1}$ cannot be optimal under belief
in $(S,E)$ (albeit perturbed). Thus, Ann and Bob must play a (perturbed)
equilibrium of matrix $M2$, which cannot discourage a deviation to $%
\widetilde{I.M2}$.

\bigskip

This is not the first time that a connection between equilibrium refinements
a la strategic stability and rationalizability is established. In signaling
games, Battigalli and Siniscalchi \cite{BS9} show that when an equilibrium
outcome satisfies the Iterated Intuitive Criterion (Cho and Kreps \cite{CK}%
), Strong-$\Delta $-Rationalizability yields a non-empty set for the
corresponding restrictions (i.e. the belief that opponents play compatibly
with the path). In \cite{C} I\ prove that Selective Rationalizability yields
the empty set for a class of non strategically stable equilibrium paths:
those that \emph{can be upset by a convincing deviation} (Osborne \cite{O}).
So, one could think that strategic stability simply requires non-emptiness
of Selective Rationalizability under the belief in the equilibrium path.
This is false. In the example above, Selective\ Rationalizability yields a
non-empty set under the belief in $(O,(S,E))$ (but does not yield $(O,(S,E))$
as unique prediction). Thus, there is no incompatibility between the belief
in the path and the rationalization of deviations based on it (unlike for
equilibrium paths that can be upset by a convincing deviation). The problem
is the incompatibility between the rationalization of deviations based on
the belief in the path and the threats that sustain the path in equilibrium.
This calls for a rationalizability procedure that takes both into account in
a given \emph{epistemic priority ordering}; in particular the "theory" that
players comply with the path will be assigned higher epistemic priority with
respect the "theory" that players implement also the equilibrium threats. In
the full version of the paper, I construct and characterize epistemically
such rationalizability procedure. The scope is expanded to an arbitrary
number of theories of opponents' behavior, of an arbitrary nature (i.e. not
just path versus full equilibrium behavior). Without the ambition to
perfectly characterize strategic stability, the application of this
rationalizability procedure to an equilibrium path and profile captures in a
general and transparent way the spirit of the strategic reasoning stories in
the background of strategic stability and related refinements.

\bigskip 

\section{Appendix}

For each $i\in I$ and $h\in H_{i}$, let $p(h)$ be the immediate predecessor
of $h$.

\bigskip

\begin{lemma}
\label{COMPOSIZIONE}Fix a profile of rationalizable subsets of CPS's $%
(\Delta _{i})_{i\in I}$.\\ For every $n\geq 0$, $i\in I$, $h\in
H_{i}(S_{i,R\Delta }^{n})\backslash H_{i}(S^{\infty })$ with $p(h)\in
H_{i}(S^{\infty })$, and $s_{i}\in S_{i}^{\infty }(h)$, there exists $%
s_{i}^{\ast }\in S_{i,R\Delta }^{n}$ such that $s_{i}^{\ast }(h^{\prime
})=s_{i}(h^{\prime })$ for all $h^{\prime }\succeq h$.
\end{lemma}

\textbf{Proof. }By $S_{i,R\Delta }^{0}=S_{i}^{\infty }$, the result trivally
holds for $n=0$. Fix $n>0$ and suppose to have proved the result for all $%
q<n $. Fix $i\in I$, $h\in H_{i}(S_{i,R\Delta }^{n})\backslash
H_{i}(S^{\infty }) $ with $p(h)\in H_{i}(S^{\infty })$, and $s_{i}\in
S_{i}^{\infty }(h)$. Fix $\mu _{i}\in \Delta _{i}$ that satifies S2 and S3
with $\rho (\mu _{i})(h)\not=\emptyset $ (it exists by $h\in
H_{i}(S_{i,R\Delta }^{n})$) and $\mu _{i}^{\prime }$ that satisfies S3 with $%
s_{i}\in \rho (\mu _{i}^{\prime })$. For each $j\not=i$ and $s_{j}\in
S_{j}^{\infty }(h)$, letting $m:=\max \left \{ q<n:S_{j,R\Delta
}^{q}(h)\not=\emptyset \right \} $, by the Induction Hypothesis\ there
exists $s_{j}^{\ast }\in S_{j,R\Delta }^{m}(h)$ such that $s_{j}^{\ast
}(h^{\prime })=s_{j}(h^{\prime })$ for all $h^{\prime }\succeq h$. Let $\eta
_{j}^{h}(s_{j}):=s_{j}^{\ast }$.\ For each $s_{j}\in S_{j}(h)\backslash
S_{j}^{\infty }(h)$, let $\eta _{j}^{h}(s_{j}):=s_{j}$. Since $\mu _{i}$
strongly believes $S_{-i,R\Delta }^{0}=S_{-i}^{\infty }$, $h\not \in
H_{i}(S_{-i}^{\infty })$, and $p(h)\in H_{i}(S_{-i}^{\infty })$, $\mu
_{i}(S_{-i}(h)|p(h))=0$. Thus, I can construct $\mu _{i}^{\ast }$ that
satisfies S2 and S3 as (i) $\mu _{i}^{\ast }(\cdot |h^{\prime })=\mu
_{i}(\cdot |h^{\prime })$ for all $h^{\prime }\not \succeq h$, and (ii) $\mu
_{i}^{\ast }(s_{-i}|h^{\prime })=\mu _{i}^{\prime }(\times _{j\not=i}(\eta
_{j}^{h})^{-1}(s_{j})|h^{\prime })$ for all $h^{\prime }\succeq h$ and $%
s_{-i}=(s_{j})_{j\not=i}\in S_{-i}(h)$. By (i) and rationalizability of $%
\Delta _{i}$, $\mu _{i}^{\ast }\in $ $\Delta _{i}$. By (i) and (ii), there
exists $s_{i}^{\ast }\in \rho (\mu _{i}^{\ast })(h)\subseteq S_{i,R\Delta
}^{n}$ such that $s_{i}^{\ast }(h^{\prime })=s_{i}(h^{\prime })$ for all $%
h^{\prime }\succeq h$. $\blacksquare $

\bigskip

\textbf{Proof of Proposition \ref{NO S3}. }Fix $n\in 
%TCIMACRO{\U{2115} }%
%BeginExpansion
\mathbb{N}
%EndExpansion
$, $i\in I$, $\mu _{i}\in \Delta _{i}$ that satifies S2 at $n$, and $%
s_{i}\in \rho (\mu _{i})\cap S_{i}^{\infty }$. Fix $\mu _{i}^{\prime }$ that
satisfies S3 with $s_{i}\in \rho (\mu _{i}^{\prime })$. Fix $h\in
H_{i}(s_{i})\backslash H_{i}(S^{\infty })$ with $p(h)\in H_{i}(S^{\infty })$%
. For each $j\not=i$ and $s_{j}\in S_{j}^{\infty }(h)$, letting $m:=\max
\left \{ q<n:S_{j,R\Delta }^{q}(h)\not=\emptyset \right \} $, by Lemma \ref%
{COMPOSIZIONE}\ there exists $s_{j}^{\ast }\in S_{j,R\Delta }^{m}(h)$ such
that $s_{j}^{\ast }(h^{\prime })=s_{j}(h^{\prime })$ for all $h^{\prime
}\succeq h$. Let $\eta _{j}^{h}(s_{j}):=s_{j}^{\ast }$.\ For each $s_{j}\in
S_{j}(h)\backslash S_{j}^{\infty }(h)$, let $\eta _{j}^{h}(s_{j}):=s_{j}$.
Since $\mu _{i}$ strongly believes $S_{-i,R\Delta }^{0}=S_{-i}^{\infty }$, $%
h\not \in H_{i}(S_{-i}^{\infty })$, and $p(h)\in H_{i}(S_{-i}^{\infty })$, $%
\mu _{i}(S_{-i}(h)|p(h))=0$. Thus, there exists $\mu _{i}^{\ast }$ that
satisfies S2 and S3 such that (i) $\mu _{i}^{\ast }(\cdot |h)=\mu _{i}(\cdot
|h)$ for all $h\in H_{i}(S^{\infty })$, and (ii) $\mu _{i}^{\ast
}(s_{-i}|h^{\prime })=\mu _{i}^{\prime }(\times _{j\not=i}(\eta
_{j}^{h})^{-1}(s_{j})|h^{\prime })$ for all $h\in H_{i}(s_{i})\backslash
H_{i}(S^{\infty })$ with $p(h)\in H_{i}(S^{\infty })$, $h^{\prime }\succeq h$%
, and $s_{-i}=(s_{j})_{j\not=i}\in S_{-i}(h)$. By (i) and rationalizability
of $\Delta _{i}$, $\mu _{i}^{\ast }\in $ $\Delta _{i}$. By (i) and (ii), $%
s_{i}\in \rho (\mu _{i})\subseteq S_{i,R\Delta }^{n}$. $\blacksquare $

\bigskip

\textbf{Proof of Proposition \ref{NO NEED RESTRICT}. }For each $i\in I$, let 
$\overline{\Delta }_{i}$ be the set of all $\mu _{i}\in \Delta _{i}$ that
satisfy S3 and S2 under $(\Delta _{j})_{j\in I}$ for all $n\in 
%TCIMACRO{\U{2115} }%
%BeginExpansion
\mathbb{N}
%EndExpansion
$. By finiteness,\footnote{%
Or other milder conditions which guarantee that every $s_{i}\in S_{i,R\Delta
}^{\infty }$ is a sequential best reply to some belief $\mu _{i}$ that
strongly believes $((S_{j,R\Delta }^{q})_{j\in I})_{q=0}^{\infty }$.} (1) $%
S_{R\overline{\Delta }}^{\infty }=S_{R\overline{\Delta }}^{1}=\times _{i\in
I}\rho (\overline{\Delta }_{i})=S_{R\Delta }^{\infty }$. Let $\mu _{i}\in
\Delta _{i}^{\ast }$ if and only if there exists $\overline{\mu }_{i}\in 
\overline{\Delta }_{i}$ such that $\mu _{i}(\cdot |h)=\overline{\mu }%
_{i}(\cdot |h)$ for all $h\in H_{i}(S^{\infty })$. Obviously, $\overline{%
\Delta }_{i}\subseteq \Delta _{i}^{\ast }$ for all $i\in I$; thus, $S_{R%
\overline{\Delta }}^{1}\subseteq S_{R\Delta ^{\ast }}^{1}$.

I show first that $\Delta _{i}^{\ast }$ is rationalizable. Fix $\mu _{i}$
and $\overline{\mu }_{i}\in \Delta _{i}^{\ast }$ such that $\mu _{i}(\cdot
|h)=\overline{\mu }_{i}(\cdot |h)$ for all $h\in H_{i}(S^{\infty })$. Then,
there exists $\overline{\overline{\mu }}_{i}\in \overline{\Delta }_{i}$ such
that $\overline{\overline{\mu }}_{i}(\cdot |h)=\overline{\mu }_{i}(\cdot
|h)=\mu _{i}(\cdot |h)$ for all $h\in H_{i}(S^{\infty })$, so $\mu _{i}\in
\Delta _{i}^{\ast }$.

Fix $i\in I$ and $s_{i}\in S_{i,R\Delta ^{\ast }}^{1}$.\ Fix $\mu _{i}\in
\Delta _{i}^{\ast }$ such that $s_{i}\in \rho (\mu _{i})$. By definition of $%
\Delta _{i}^{\ast }$, there exists $\overline{\mu }_{i}\in \overline{\Delta }%
_{i}\subseteq \Delta _{i}^{\ast }$ such that $\mu _{i}(\cdot |h)=\overline{%
\mu }_{i}(\cdot |h)$ for all $h\in H_{i}(S^{\infty })$. Thus, there exists $%
\overline{s}_{i}\in \rho (\overline{\mu }_{i})\subseteq \rho (\overline{%
\Delta }_{i})=S_{i,R\overline{\Delta }}^{1}$ such that $\overline{s}%
_{i}(h)=s_{i}(h)$ for all $h\in H_{i}(S^{\infty })\supseteq H_{i}\left(
S_{R\Delta ^{\ast }}^{1}\right) $. Hence, by $S_{R\overline{\Delta }%
}^{1}\subseteq S_{R\Delta ^{\ast }}^{1}$, ($\bigstar $) $H_{i}(S_{j,R\Delta
^{\ast }}^{1})\cap H_{i}(S^{\infty })=H_{i}(S_{j,R\overline{\Delta }%
}^{1})\cap H_{i}(S^{\infty })$ for all $j\not=i$, and by $\zeta (S_{R\Delta
^{\ast }}^{1})\subseteq \zeta (S^{\infty })$, (2) $\zeta (S_{R\Delta ^{\ast
}}^{1})=\zeta (S_{R\overline{\Delta }}^{1})$.

Since $\overline{s}_{i}\in S_{i,R\overline{\Delta }}^{2}=S_{i,R\overline{%
\Delta }}^{1}$, there exists $\overline{\mu }_{i}\in \overline{\Delta }_{i}$
that strongly believes $(S_{j,R\overline{\Delta }}^{1})_{j\not=i}$ such that 
$\overline{s}_{i}\in \rho (\overline{\mu }_{i})$. For each $h\in
H_{i}(S^{\infty })$ and $s_{-i}\in S_{-i}$ with $\overline{\mu }%
_{i}(s_{-i}|h)>0$, since $\overline{\mu }_{i}$ strongly believes in $%
S_{-i}^{\infty }$, $\overline{s}_{i}\in S_{i}^{\infty }$, and $\overline{s}%
_{i}(h^{\prime })=s_{i}(h^{\prime })$ for all $h^{\prime }\in
H_{i}(S^{\infty })$, $\zeta (\overline{s}_{i},s_{-i})=\zeta (s_{i},s_{-i})$.
Thus, by $\overline{s}_{i}\in \rho (\overline{\mu }_{i})$, $s_{i}$ is a
continuation best reply to $\overline{\mu }_{i}(\cdot |h)$ too. Fix $\mu
_{i}^{\prime }$ that satisfies S3 such that $s_{i}\in \rho (\mu _{i}^{\prime
})$. Fix $h\in H_{i}(s_{i})\backslash H_{i}(S^{\infty })$ with $p(h)\in
H_{i}(S^{\infty })$. For each $j\not=i$ and $s_{j}\in S_{j}^{\infty }(h)$,
letting $m:=\max \left \{ q=0,1:S_{j,R\Delta ^{\ast }}^{q}(h)\not=\emptyset
\right \} $, by Lemma \ref{COMPOSIZIONE} there exists $s_{j}^{\ast }\in
S_{j,R\Delta ^{\ast }}^{m}(h)$ such that $s_{j}^{\ast }(h^{\prime
})=s_{j}^{\prime }(h^{\prime })$ for all $h^{\prime }\succeq h$. Let $\eta
_{j}^{h}(s_{j}):=s_{j}^{\ast }$.\ For each $s_{j}\in S_{j}(h)\backslash
S_{j}^{\infty }(h)$, let $\eta _{j}^{h}(s_{j}):=s_{j}$. Since $\overline{\mu 
}_{i}$ strongly believes $S_{-i}^{\infty }$, $h\not \in H_{i}(S_{-i}^{\infty
})$, and $p(h)\in H_{i}(S_{-i}^{\infty })$, $\overline{\mu }%
_{i}(S_{-i}(h)|p(h))=0$. For all $j\not=i$, by $S_{R\overline{\Delta }%
}^{1}\subseteq S_{R\Delta ^{\ast }}^{1}$ and ($\bigstar $), $\overline{\mu }%
_{i}(S_{j,R\Delta ^{\ast }}^{1}\times S_{-i,j}|h)=1$ for all $h\in
H_{i}(S_{j,R\Delta ^{\ast }}^{1})\cap H_{i}(S^{\infty })$. Thus, there
exists $\mu _{i}^{\ast }$ that satisfies S3 and strongly believes $%
(S_{j,R\Delta ^{\ast }}^{1})_{j\not=i}$ such that (i) $\mu _{i}^{\ast
}(\cdot |h)=\overline{\mu }_{i}(\cdot |h)$ for all $h\in H_{i}(S^{\infty })$%
, and (ii) $\mu _{i}^{\ast }(s_{-i}|h^{\prime })=\mu _{i}^{\prime }(\times
_{j\not=i}(\eta _{j}^{h})^{-1}(s_{j})|h^{\prime })$ for all $h\in
H_{i}(s_{i})\backslash H_{i}(S^{\infty })$ with $p(h)\in H_{i}(S^{\infty })$%
, $h^{\prime }\succeq h$, and $s_{-i}=(s_{j})_{j\not=i}\in S_{-i}(h)$. By
(i) and rationalizability of $\Delta _{i}^{\ast }$, $\mu _{i}^{\ast }\in
\Delta _{i}^{\ast }$. By (i) and (ii), $s_{i}\in \rho (\mu _{i})\subseteq
S_{i,R\Delta ^{\ast }}^{2}$. Thus, (3) $S_{R\Delta ^{\ast }}^{\infty
}=S_{R\Delta ^{\ast }}^{1}$. By 1-2-3, $\zeta (S_{R\Delta ^{\ast }}^{\infty
})=\zeta (S_{R\Delta }^{\infty })$. $\blacksquare $

\bigskip

\bigskip

\textbf{PROOF OF THEOREM \ref{T4}.}

\bigskip

First, I prove a generalized version of Theorem \ref{T4}. Applying this
generalized version to Rationalizability yields the hypotheses to apply it
to Selective Rationalizability and prove Theorem \ref{T4}.

Consider this generalized rationalizability procedure:

\begin{definition}
Fix a profile of compact subsets of CPS's $(\Delta _{i})_{i\in I}$. Fix
another profile of compact subsets of CPS's $(\Delta _{i}^{G})_{i\in I}$.
Fix $n\geq 1$ and, if $n>1$, suppose to have defined $((S_{i,G}^{q})_{i\in
I})_{q=1}^{n-1}$. For every $i\in I$ and $s_{i}\in S_{i}$, let $s_{i}\in
S_{i,G}^{n}$ if and only if there exists $\mu _{i}\in \Delta _{i}$ such
that:\medskip

\begin{enumerate}
\item[G1] $s_{i}\in \rho (\mu _{i})$;

\item[G2] $\mu _{i}$ strongly believes $S_{j,G}^{q}$ for all $j\not=i$ and $%
q<n$;

\item[G3] $\mu _{i}\in \Delta _{i}^{G}$.
\end{enumerate}

Call $\Delta _{i}^{n,G}$ the set of all $\mu _{i}\in \Delta _{i}$ that
satisfy G2 and G3.\medskip

Finally, let $S_{i,G}^{\infty }=\cap _{n\geq 1}S_{i,G}^{n}$ and $\Delta
_{i}^{\infty ,G}=\cap _{n\geq 1}\Delta _{i}^{n,G}$.
\end{definition}

Consider the following property for a Cartesian event $E=\times _{i\in
I}E_{i}\subseteq \Omega $.

\begin{definition}
A Cartesian event $E=\times _{i\in I}E_{i}$ satisfies the "completeness
property" if for every $i\in I$, $t_{i}\in $\emph{$\mathrm{Proj}$}$%
_{T_{i}}E_{i}$, $s_{i}\in \rho (f_{i}(t_{i}))$, and maps\footnote{%
Note that the maps are injective.} $(\tau _{j})_{j\not=i}$ with $\tau _{j}:%
\overline{s}_{j}\in $\emph{$\mathrm{Proj}$}$_{S_{j}}E_{j}\mapsto (\overline{s%
}_{j},t_{j})\in E_{j}$ for all $j\not=i$, there exists $t_{i}^{\prime }\in
T_{i}$ such that $(s_{i},t_{i}^{\prime })\in E_{i}$, $f_{i}(t_{i}^{\prime
})=f_{i}(t_{i})$, and $g_{i,h}(t_{i}^{\prime })\left[ \tau _{j}(s_{j})\times
\Omega _{-i,j}\right] =f_{i,h}(t_{i})\left[ s_{j}\times S_{-i,j}\right] $
for all $h\in H_{i}$, $j\not=i$, and $s_{j}\in $\emph{$\mathrm{Proj}$}$%
_{S_{j}}E_{j}$.
\end{definition}

Now I can state a generalized characterization theorem.\footnote{%
The event $E$ can be empty, just like $CSB^{\infty }(R)\cap \lbrack \Delta ]$
in Theorem \ref{T4}.}

\begin{lemma}
\label{L2}Fix a closed, Cartesian event $E=\times _{i\in I}E_{i}\subseteq R$
with the completeness property such that for each $i\in I$, $f_{i}(\mathrm{%
Proj}_{T_{i}}E)=\Delta _{i}\cap \Delta _{i}^{G}$ (which implies $S_{G}^{1}=$%
\emph{$\mathrm{Proj}$}$_{S}E$).\footnote{$\subseteq $ is guaranteed by the
completeness property of $E$; $\supseteq $ is guaranteed by the fact that $%
E\subseteq R$.}

\noindent
Then, for every $n\in 
%TCIMACRO{\U{2115} }%
%BeginExpansion
\mathbb{N}
%EndExpansion
$, $CSB^{n-1}(E)$ has the completeness property and for each $i\in I$, $%
f_{i}(\mathrm{Proj}_{T_{i}}CSB^{n-1}(E))\\=\Delta _{i}^{n,G}$ if $n=1$, and $%
f_{i}(\mathrm{Proj}_{T_{i}}CSB_{i}(CSB^{n-2}(E)))=\Delta _{i}^{n,G}$ if $n>1$
(which implies $S_{G}^{n}=$\emph{$\mathrm{Proj}$}$_{S}CSB^{n-1}(E)$).%
\footnote{%
For $n>1$, $\mathrm{Proj}_{T_{i}}CSB^{n-1}(E)$ must be substituted by $f_{i}(%
\mathrm{Proj}_{T_{i}}CSB_{i}(CSB^{n-2}(E)))$ because for some $j\not=i$, $%
CSB_{j}(CSB^{n-2}(E))$ may be empty (and thus $CSB^{n-1}(E)$ too) while $%
\Delta _{i}^{n,G}$ is not.}

Moreover, $CSB^{\infty }(E)$ has the completeness property and for each $%
i\in I$, $f_{i}(\mathrm{Proj}_{T_{i}}CSB^{\infty }(E))=\Delta _{i}^{\infty
,G}$ (which implies $S_{G}^{\infty }=$\emph{$\mathrm{Proj}$}$_{S}CSB^{\infty
}(E)$).\footnote{%
Finiteness implies that for every $s_{i}\in S_{i,G}^{\infty }$, $s_{i}\in
\rho (\mu _{i})$ for some $\mu _{i}\in \Delta _{i}^{\infty ,G}$, but it can
be substituted by mild regularity conditions (see, for instance, Battigalli 
\cite{B2}).}
\end{lemma}

\textbf{Proof. }For finite $n$, the proof is by induction.

\bigskip

\textbf{Induction Hypothesis (n=1,...,m): }the Lemma holds for $n=1,...m$.

\bigskip

\textbf{Basis step (n=1): }the Lemma holds for $n=1$ by hypothesis.

\bigskip

\textbf{Inductive step (n=m+1): }Let $F=\times _{i\in I}F_{i}:=CSB^{m-1}(E)$
and $G=\times _{i\in I}G_{i}:=CSB^{m}(E)$, where for all $i\in I$, $%
F_{i}=CSB_{i}(CSB^{m-2}(E))$ and $G_{i}=CSB_{i}(F)$

\bigskip

Fix $i\in I$ and $\mu _{i}\in \Delta _{i}^{m+1,G}\subseteq \Delta _{i}^{m,G}$%
. Then, by the Induction Hypothesis, there exists $t_{i}\in $\emph{$\mathrm{%
Proj}$}$_{T_{i}}F_{i}$ such that $f_{i}(t_{i})=\mu _{i}$. Fix maps $(\tau
_{j})_{j\not=i}$ with $\tau _{j}:\overline{s}_{j}\in $\emph{$\mathrm{Proj}$}$%
_{S_{j}}F_{j}\mapsto (\overline{s}_{j},t_{j})\in F_{j}$ for all $j\not=i$.
By the Induction Hypothesis, $F$ has the completeness property. So, there
exists $(s_{i}^{\prime },t_{i}^{\prime })\in F_{i}$ such that $%
f_{i}(t_{i}^{\prime })=f_{i}(t_{i})=\mu _{i}$, and for every $h\in H_{i}$, $%
j\not=i$, and $s_{j}\in $\emph{$\mathrm{Proj}$}$_{S_{j}}F_{j}$, $%
g_{i,h}(t_{i}^{\prime })\left[ \tau _{j}(s_{j})\times \Omega _{-i,j}\right]
=f_{i,h}(t_{i})\left[ s_{j}\times S_{-i,j}\right] $. Then, since $%
f_{i}(t_{i})=\mu _{i}\in \Delta _{i}^{m+1,G}$ strongly believes $%
S_{j,G}^{m}= $\emph{$\mathrm{Proj}$}$_{S_{j}}F_{j}$ (by the Induction
Hypothesis), $g_{i}(t_{i}^{\prime })$ strongly believes $F_{j}$. So, $%
(s_{i}^{\prime },t_{i}^{\prime })\in SB_{i}(F_{-i})\cap F_{i}$. Thus, $%
(s_{i}^{\prime },t_{i}^{\prime })\in G_{i}$.

Fix $i\in I$ and $t_{i}\in $\emph{$\mathrm{Proj}$}$_{T_{i}}G_{i}$. Since $%
t_{i}\in $\emph{$\mathrm{Proj}$}$_{T_{i}}F_{i}$, by the Induction Hypothesis 
$f_{i}(t_{i})\in \Delta _{i}^{m,G}$. Since $t_{i}\in $\emph{$\mathrm{Proj}$}$%
_{T_{i}}SB_{i}(F)$, $g_{i}(t_{i})$ strongly believes $F_{j}$ for all $%
j\not=i $, hence $f_{i}(t_{i})$ strongly believes \emph{$\mathrm{Proj}$}$%
_{S_{j}}F_{j}$. By the Induction Hypothesis \emph{$\mathrm{Proj}$}$%
_{S_{j}}F_{j}=S_{j}^{m}$. So $f_{i}(t_{i})\in \Delta _{i}^{m+1,G}$.

\bigskip

Now I show that $G$ has the completeness property. Fix $i\in I$, $t_{i}\in $%
\emph{$\mathrm{Proj}$}$_{T_{i}}G_{i}\subseteq $\emph{$\mathrm{Proj}$}$%
_{T_{i}}F_{i}$, $s_{i}\in \rho (f_{i}(t_{i}))$, and maps $(\tau
_{j})_{j\not=i}$ with $\tau _{j}:\overline{s}_{j}\in \mathrm{Proj}%
_{S_{j}}G_{j}\mapsto (\overline{s}_{j},t_{j})\in G_{j}\subseteq F_{j}$ for
all $j\not=i$. Extend each $\tau _{j}$ to $\tau _{j}^{\prime }:\overline{s}%
_{j}\in $\emph{$\mathrm{Proj}$}$_{S_{j}}F_{j}\mapsto (\overline{s}%
_{j},t_{j})\in F_{j}$ in such a way that for every $s_{j}\in $\emph{$\mathrm{%
Proj}$}$_{S_{j}}G_{j}$, $\tau _{j}^{\prime }(s_{j})=\tau _{j}(s_{j})$. By
the Induction Hypothesis, $F$ has the completeness property. So, there
exists $t_{i}^{\prime }\in T_{i}$ such that $(s_{i},t_{i}^{\prime })\in
F_{i} $, $f_{i}(t_{i}^{\prime })=f_{i}(t_{i})$, and for every $h\in H_{i}$, $%
j\not=i$, and $s_{j}\in $\emph{$\mathrm{Proj}$}$_{S_{j}}F_{j}\supseteq $%
\emph{$\mathrm{Proj}$}$_{S_{j}}G_{j}$, $g_{i,h}(t_{i}^{\prime })[\tau
_{j}^{\prime }(s_{j})\times \Omega _{-i,j}]=f_{i,h}(t_{i})\left[ s_{j}\times
S_{-i,j}\right] $. Since $t_{i}\in $\emph{$\mathrm{Proj}$}$%
_{T_{i}}SB_{i}(F_{-i})$, $f_{i}(t_{i})$ strongly believes \emph{$\mathrm{Proj%
}$}$_{S_{j}}F_{j}$ for all $j\not=i$. Then, by construction, $%
g_{i}(t_{i}^{\prime })$ strongly believes $F_{j}$. So $(s_{i},t_{i}^{\prime
})\in SB_{i}(F_{-i})$. Thus, $(s_{i},t_{i}^{\prime })\in G_{i}$. Note that
for every $h\in H_{i}$, $j\not=i$, and $s_{j}\in $\emph{$\mathrm{Proj}$}$%
_{S_{j}}G_{j}$, $g_{i,h}(t_{i}^{\prime })[\tau _{j}(s_{j})\times \Omega
_{-i,j}]=f_{i,h}(t_{i})\left[ s_{j}\times S_{-i,j}\right] $. $\square $

\bigskip

Now I prove that the lemma holds for $n=\infty $. By finiteness, there is $%
M\in 
%TCIMACRO{\U{2115} }%
%BeginExpansion
\mathbb{N}
%EndExpansion
$ such that $S_{G}^{\infty }=S_{G}^{M}$. Let $F:=CSB^{M}(E)$ and $%
G:=CSB^{\infty }(E)=\cap _{n\geq 0}CSB^{n}(E)$.

\bigskip

Fix $i\in I$ and $\mu _{i}\in \Delta _{i}^{\infty ,G}=\Delta _{i}^{M+1,G}$.

By finiteness, the existence of such $\mu _{i}$ implies that $\Delta
_{j}^{\infty ,G}\not=\emptyset $ for all $j\not=i$, so $S_{G}^{\infty
}\not=\emptyset $. As shown above, for every $s\in S_{G}^{\infty }$ and $%
q\geq 0$, $(\left \{ s\right \} \times T)\cap CSB^{q}(E)\not=\emptyset $.
Since each $CSB^{q}(E)$ and $(\left \{ s\right \} \times T)$ are closed (see
Section 4), $((\left \{ s\right \} \times T)\cap CSB^{q}(E))_{q\geq 0}$ is a
sequence of nested, nonempty closed sets, so it has the finite intersection
property. Since $\Omega $ is compact, $(\left \{ s\right \} \times T)\cap
G\not=\emptyset $. Then, for each $j\in I$, there exists $\tau _{j}:%
\overline{s}_{j}\in $\emph{$\mathrm{Proj}$}$_{S_{j}}F\mapsto (\overline{s}%
_{j},t_{j})\in \mathrm{Proj}_{\Omega _{j}}G$.

As shown above, there exists $t_{i}\in $\emph{$\mathrm{Proj}$}$_{T_{i}}F$
such that $f_{i}(t_{i})=\mu _{i}$, and $F$ has the completeness property.
So, there exists $(s_{i}^{\prime },t_{i}^{\prime })\in \mathrm{Proj}_{\Omega
_{i}}F$ such that for every $i\in I$, $f_{i}(t_{i}^{\prime })=f_{i}(t_{i})$,
and for every $h\in H_{i}$, $j\not=i$, and $s_{j}\in $\emph{$\mathrm{Proj}$}$%
_{S_{j}}F$, $g_{i,h}(t_{i}^{\prime })[\tau _{j}(s_{j})\times \Omega
_{-i,j}]=f_{i,h}(t_{i})\left[ s_{j}\times S_{-i,j}\right] $. Then, since $%
f_{i}(t_{i})=\mu _{i}\in \Delta _{i}^{\infty ,G}$ strongly believes $%
S_{j,G}^{\infty }=$\emph{$\mathrm{Proj}$}$_{S_{j}}F$ (shown above), $%
g_{i}(t_{i}^{\prime })$ strongly believes $\mathrm{Proj}_{\Omega _{j}}G$.
Hence, for each $q\geq M$, since $S_{G}^{q}=S_{G}^{\infty }$ and $%
CSB^{q}(E)\supset G$, $g_{i}(t_{i}^{\prime })$ strongly believes \emph{$%
\mathrm{Proj}$}$_{\Omega _{j}}CSB^{q}(E)$. So, $(s_{i}^{\prime
},t_{i}^{\prime })\in SB_{i}(CSB^{q}(E))$. Repeating for each $i\in I$, $%
CSB^{q+1}(E)\not=\emptyset $. Then $(s_{i}^{\prime },t_{i}^{\prime })\in 
\mathrm{Proj}_{\Omega _{i}}CSB^{q+1}(E)$ for all $q\geq M$. Thus $%
(s_{i}^{\prime },t_{i}^{\prime })\in \mathrm{Proj}_{\Omega
_{i}}G\not=\emptyset $.

Fix $i\in I$ and $t_{i}\in $\emph{$\mathrm{Proj}$}$_{T_{i}}G$. For every $%
q\geq 1$, $t_{i}\in $\emph{$\mathrm{Proj}$}$_{T_{i}}CSB^{q-1}(E)$, thus, as
shown above, $f_{i}(t_{i})\in \Delta _{i}^{q,G}$. Then, $f_{i}(t_{i})\in
\Delta _{i}^{\infty ,G}$.

\bigskip

Now I show that $G$ has the completeness property. Fix $i\in I$, $t_{i}\in $%
\emph{$\mathrm{Proj}$}$_{T_{i}}G\subseteq $\emph{$\mathrm{Proj}$}$_{T_{i}}F$%
, $s_{i}\in \rho (f_{i}(t_{i}))$, and maps $(\tau _{j})_{j\not=i}$ with $%
\tau _{j}:\overline{s}_{j}\in \mathrm{Proj}_{S_{j}}G\mapsto (\overline{s}%
_{j},t_{j})\in \mathrm{Proj}_{\Omega _{j}}G\subseteq \mathrm{Proj}_{\Omega
_{j}}F$ for all $j\not=i$. As shown above, \emph{$\mathrm{Proj}$}$%
_{S}G=S_{G}^{\infty }=S_{G}^{M}=$\emph{$\mathrm{Proj}$}$_{S}F$, and $F$ has
the completeness property. So, there exists $t_{i}^{\prime }\in T_{i}$ such
that $(s_{i},t_{i}^{\prime })\in \mathrm{Proj}_{\Omega _{i}}F$, $%
f_{i}(t_{i}^{\prime })=f_{i}(t_{i})$, and for every $h\in H_{i}$, $j\not=i$,
and $s_{j}\in $\emph{$\mathrm{Proj}$}$_{S_{j}}F=$\emph{$\mathrm{Proj}$}$%
_{S_{j}}G$, $g_{i,h}(t_{i}^{\prime })[\tau _{j}(s_{j})\times \Omega
_{-i,j}]=f_{i,h}(t_{i})\left[ s_{j}\times S_{-i,j}\right] $. Since $t_{i}\in 
$\emph{$\mathrm{Proj}$}$_{T_{i}}SB_{i}(F)$, $f_{i}(t_{i})$ strongly believes 
\emph{$\mathrm{Proj}$}$_{S_{j}}F=$\emph{$\mathrm{Proj}$}$_{S_{j}}G$ for all $%
j\not=i$. Then, $g_{i}(t_{i}^{\prime })$ strongly believes $\mathrm{Proj}%
_{\Omega _{j}}G$. Hence, for each $q\geq M$, since \emph{$\mathrm{Proj}$}$%
_{S}F=\mathrm{Proj}_{S}CSB^{q}(E)=$\emph{$\mathrm{Proj}$}$_{S}G$ and $%
CSB^{q}(E)\supset G$, $g_{i}(t_{i}^{\prime })$ strongly believes \emph{$%
\mathrm{Proj}$}$_{\Omega _{j}}CSB^{q}(E)$. So, $(s_{i},t_{i}^{\prime })\in
SB_{i}(CSB^{q}(E))$. Then $(s_{i},t_{i}^{\prime })\in \mathrm{Proj}_{\Omega
_{i}}CSB^{q+1}(E)$ for all $q\geq M$. Thus $(s_{i},t_{i}^{\prime })\in 
\mathrm{Proj}_{\Omega _{i}}G$. $\blacksquare $

\bigskip

\textbf{Proof of Theorem \ref{T4}.}

\bigskip

For each $i\in I$, let $\Delta _{i}^{\infty ,G}$ be the set of CPS's that
satisfy S3. Theorem \ref{T4} is given by Lemma \ref{L2} with $E=\left[
\Delta \right] \cap CSB^{\infty }(R)$ and $\Delta _{i}^{G}=\Delta
_{i}^{\infty ,G}$ for all $i\in I$. I show that $\left[ \Delta \right] \cap
CSB^{\infty }(R)$ satisfies the hypotheses of Lemma \ref{L2}.

\bigskip

The event $R=\times _{i\in I}R_{i}$ is closed (see Section 4). Now I\ show
that it has the completeness property. Fix $i\in I$, $t_{i}\in $\emph{$%
\mathrm{Proj}$}$_{T_{i}}R$, $s_{i}\in \rho (f_{i}(t_{i}))$, and, for each $%
j\not=i$, $\tau _{j}:\overline{s}_{j}\in $\emph{$\mathrm{Proj}$}$%
_{S_{j}}R\mapsto (\overline{s}_{j},t_{j})\in R_{j}$. Extend each $\tau _{j}$
to $\tau _{j}^{\prime }:\overline{s}_{j}\in S_{j}\mapsto (\overline{s}%
_{j},t_{j})\in \Omega _{j}$ in such a way that for every $s_{j}\in $\emph{$%
\mathrm{Proj}$}$_{S_{j}}R$, $\tau _{j}^{\prime }(s_{j})=\tau _{j}(s_{j})$.
Define $\nu _{i}\in (\Delta (S_{-i}\times T_{-i}))^{H_{i}}$ as $\nu
_{i}(\times _{j\not=i}\tau _{j}^{\prime }(s_{j})|h)=f_{i,h}(t_{i})[s_{-i}]$
for all $h\in H_{i}$ and $s_{-i}=(s_{j})_{j\not=i}\in S_{-i}$ (it is well
defined because each $\tau _{j}^{\prime }$ is injective). It is easy to
verify that $\nu _{i}$ is a CPS given that $f_{i}(t_{i})$ is a CPS.\footnote{%
A detailed argument for this under finiteness can be found in \cite{BP}, in
the proof of Lemma 1.} By ontoness of $g_{i}$, there exists $t_{i}^{\prime
}\in T_{i}$ such that $g_{i}(t_{i}^{\prime })=\nu _{i}$. Clearly, $%
f_{i}(t_{i}^{\prime })=f_{i}(t_{i})$, which implies $(s_{i},t_{i}^{\prime
})\in R_{i}$, and $g_{i,h}(t_{i}^{\prime })\left[ \tau _{j}(s_{j})\times
\Omega _{-i,j}\right] =f_{i,h}(t_{i})\left[ s_{j}\times S_{-i,j}\right] $
for all $h\in H_{i}$, $j\not=i$, and $s_{j}\in $\emph{$\mathrm{Proj}$}$%
_{S_{j}}R$.

Note that (trivially) $f_{i}(\mathrm{Proj}_{T_{i}}R_{i})=\Delta
^{H_{i}}(S_{-i})$.

So, I can apply Lemma \ref{L2} with $E=R$ and $\Delta _{i}^{G}=\Delta
^{H_{i}}(S_{-i})$, so that $((S_{i,G}^{n})_{i\in I})_{n=0}^{\infty }$ is
Rationalizability. Thus, $CSB^{\infty }(R)$ is a closed Cartesian event with
the completeness property where $f_{i}(\mathrm{Proj}_{T_{i}}CSB^{\infty
}(R))=\Delta _{i}^{\infty ,G}$ for all $i\in I$. Then, it is easy to check
that $E=\left[ \Delta \right] \cap CSB^{\infty }(R)$ is a closed Cartesian
event with the completeness property where $f_{i}(\mathrm{Proj}%
_{T_{i}}E)=\Delta _{i}\cap \Delta _{i}^{\infty ,G}$ for all $i\in I$. $%
\blacksquare $

%\nocite{*}
\bibliographystyle{eptcs}
\bibliography{mainpaper}
\end{document}